\documentclass[sigconf]{acmart}
\usepackage{multirow}
\usepackage{makecell}
\usepackage{pifont}

\AtBeginDocument{%
  }

\setcopyright{acmcopyright}
\copyrightyear{2023}
\acmYear{2023}
\acmDOI{XXXXXXX.XXXXXXX}

\acmConference[xx]{xx}{xx}{xx}
\acmPrice{15.00}
\acmISBN{978-1-4503-XXXX-X/18/06}

\begin{document}

\title{Self-Supervised Multi-Modal Sequential Recommendation}


\author{Kunzhe Song}
\email{kz_song@stu.pku.edu.cn}
\affiliation{
  \institution{Peking University}
  \city{Beijing}
  \country{China}}

\author{Qingfeng Sun}
\email{qins@microsoft.com}
\affiliation{
  \institution{Microsoft}
  \city{Beijing}
  \country{China}}

\author{Can Xu}
\email{caxu@microsoft.com}
\affiliation{
  \institution{Microsoft}
  \city{Beijing}
  \country{China}}
  
\author{Kai Zheng}
\email{zhengkai@microsoft.com}
\affiliation{
\institution{Microsoft}
\city{Beijing}
\country{China}}

\author{Yaming Yang}
\email{yang.yaming@microsoft.com}
\affiliation{
  \institution{Microsoft}
  \city{Beijing}
  \country{China}}

\begin{abstract}
With the increasing development of e-commerce and online services, personalized recommendation systems have become crucial for enhancing user satisfaction and driving business revenue. Traditional sequential recommendation methods that rely on explicit item IDs encounter challenges in handling item cold start and domain transfer problems. Recent approaches have attempted to use modal features associated with items as a replacement for item IDs, enabling the transfer of learned knowledge across different datasets. However, these methods typically calculate the correlation between the model's output and item embeddings, which may suffer from inconsistencies between high-level feature vectors and low-level feature embeddings, thereby hindering further model learning. 
To address this issue, we propose a dual-tower retrieval architecture for sequence recommendation. In this architecture, the predicted embedding from the user encoder is used to retrieve the generated embedding from the item encoder, thereby alleviating the issue of inconsistent feature levels. Moreover, in order to further improve the retrieval performance of the model, we also propose a self-supervised multi-modal pretraining method inspired by the consistency property of contrastive learning. This pretraining method enables the model to align various feature combinations of items, thereby effectively generalizing to diverse datasets with different item features. 
We evaluate the proposed method on five publicly available datasets and conduct extensive experiments. The results demonstrate significant performance improvement of our method. The code and pre-trained model for our method are publicly available at \href{https://github.com/kz-song/MMSRec}{https://github.com/kz-song/MMSRec}.

\end{abstract}



\maketitle

\section{Introduction}

In recent years, recommendation systems have become an essential component of many online services and e-commerce platforms. Unlike when searching for products with a specific intent, users browsing online content often lack a clear intention. As it is impractical to display all available items on a platform, personalized recommendation systems have emerged. These systems \cite{cheng2016wide} aim to provide users with customized recommendations that help them discover new items or products of interest. Personalized recommendation systems focus on mining the similarity between users and items by obtaining the probability distribution of interactions between users and all items. Early recommendation methods, such as collaborative filtering \cite{rendle2012bpr}, modeled users and items in a hidden space as vectors based on their past interaction information. Later, deep learning-based recommendation methods modeled users using deep neural networks, introducing user features \cite{covington2016deep}, multi-interest \cite{cen2020controllable, li2022improving}, and other factors to more accurately predict user preferences. However, these methods all face certain limitation in capturing the temporal dynamics and sequence information of user behavior.

Sequential recommendation methods \cite{hidasi2015session, tang2018personalized} have emerged as a promising approach to modeling user behavior in a more fine-grained and dynamic way. These methods exploit the temporal order of user interactions to predict the next item that a user is likely to engage with. By considering the sequential dependencies of user behavior, these methods can also provide more accurate and diverse recommendations, especially for long-term user preferences.
In the early days, simple sequential models like Markov Chains \cite{rendle2010factorizing, he2016fusing} were employed to capture sequential patterns in user behavior data. However, this approach faced limitations in capturing long-term dependencies and handling variable-length sequences. 
With the advent of deep learning techniques, more advanced methods have been proposed. Recurrent Neural Networks (RNNs) \cite{cho2014learning, hidasi2015session} can capture long-term dependencies in sequential data, while Transformer-based models \cite{kang2018self, sun2019bert4rec} excel at capturing complex sequential patterns. Graph-based models \cite{wu2019session, qiu2019rethinking, chang2021sequential} can explore more complex item transition patterns in user sequences. These techniques have demonstrated superior performance compared to early methods.

Despite the potential of sequential recommendation methods, the prevalent approach is to represent items explicitly by their unique identifiers or IDs. While this approach has demonstrated efficacy in certain scenarios, it is not without limitations. Specifically, this approach is typically constrained to generating item recommendations within the same platform, as items are not readily transferable across domains, hindering cross-domain generalization of the model. Furthermore, this approach is unable to perform cold-start recommendations for items with limited interaction history on the platform, which can be considered a few-shot learning problem. Consequently, numerous methods have been proposed to address this issue. For instance, some methods mine precise cross-domain user preferences based on intra-sequence and inter-sequence item interactions\cite{yuan2020parameter, cao2022contrastive}. Other methods construct mappings that project the cold item contents to the warm item embedding space\cite{pan2022multimodal, chen2022generative}. Although these methods have made progress, they have not fully resolved the fundamental issue caused by explicitly modeling item IDs.

With the rapid advancement of pre-trained models, such as BERT \cite{devlin2018bert} and CLIP \cite{radford2021learning}, recent methods for sequential recommendation have addressed the limitations of traditional item ID indexing by leveraging item-associated modalities to enable representation transfer across diverse datasets. Pre-training on datasets with modalities facilitates efficient model transfer to other datasets with similar modalities, leading to improved performance on those datasets or even achieving zero-shot recommendation. However, existing approaches \cite{hou2022towards, ding2021zero, wang2022transrec} that employ item-associated modalities for cross-domain transfer still encounter a significant challenge that needs to be addressed: the issue of inconsistent feature levels between model outputs and item embeddings.
In most sequential recommendation methods, user interaction sequences are encoded using a sequence encoder to obtain sequence-level representations, which are then directly compared with item embeddings using cosine similarity. However, the model's output represents high-level features with respect to the sequence context, while item embeddings represent low-level features for individual items, resulting in a substantial gap in feature levels. Computing similarity directly between them would forcibly align the model's input and output, which prevent the model from generating effective representations.

To address this issue, we propose a sequential recommendation method based on a dual-tower retrieval architecture. Our approach utilizes a sequence encoder for capturing user interaction sequences and an item encoder for encoding item information, ensuring that both are represented as high-level features. Considering that user interaction sequences are fundamentally composed of items, we also share parameters between the two encoders to enable mutual reinforcement of information.
To further enhance the retrieval capability of our model, we propose a self-supervised multi-modal pre-training approach. By constructing contrastive learning tasks between each modality input and its own augment representation, as well as between representations of different modalities, our model is able to strengthen its fine-grained discrimination ability for items and learn alignment of different modality features in the latent space. This enables the pre-trained model to generalize well on downstream datasets with diverse item features.

The main contributions of this paper are summarized as follows: 
\begin{itemize}
\item {We propose a sequential recommendation method based on a dual-tower retrieval architecture. Our approach employs a sequence encoder to capture user interaction sequences and an item encoder to encode item information, addressing the issue of inconsistent feature levels between model outputs and item embeddings.}
\item{We introduce a self-supervised multi-modal pre-training approach to enhance the retrieval capability of our model. By constructing contrastive learning tasks between various modalities conbinations, our model strengthens its fine-grained discrimination ability for items and learns alignment of different modality features in the latent space.}
\item{We conduct extensive experiments on five public datasets to demonstrate the effectiveness of our proposed approach and architecture.}
\end{itemize}

\section{RELATED WORK}

\subsection{Sequential Recommendation}

Sequential recommendation has been widely investigated in recent years, and various models have been proposed to address this task. Markov Chain-based techniques have been employed to model sequential dependencies \cite{rendle2010factorizing, he2016fusing} in recommendation tasks. However, these approaches have limited capacity in capturing long-term dependencies in sequences. To overcome this limitation, recurrent neural networks (RNNs) \cite{cho2014learning}, including variants such as Long Short-Term Memory (LSTM) \cite{hochreiter1997long} and Gated Recurrent Unit (GRU) \cite{donkers2017sequential, hidasi2015session}, have been introduced for sequential recommendation. These models have demonstrated strong performance in capturing both short-term and long-term dependencies in sequences.

Moreover, due to the powerful ability of attention mechanisms, many works have incorporated attention mechanisms into RNN-based models \cite{kang2018self, sun2019bert4rec} to enhance the model's capability to attend to relevant parts of the sequence. In addition to modeling based on explicit item IDs, several works are exploring other methods to improve the model's recommendation performance. For instance, FDSA \cite{zhang2019feature} and $S^3$-Rec \cite{zhou2020s3} have introduced rich attribute information of items, while CL4SRec \cite{xie2022contrastive} and DuoRec \cite{qiu2022contrastive} have introduced self-supervised signals. However, these approaches face a common issue: they are usually limited to specific data domains or platforms due to the existence of explicit item IDs.

Recently, some studies have attempted to overcome the limitation of explicit item IDs for item cold-start and cross-domain transfer. ZESRec \cite{ding2021zero} no longer relies on item indexes but instead uses natural language descriptions as item representations. UniSRec \cite{hou2022towards} has introduced two contrastive learning tasks in the pre-training stage, enabling the model to learn universal item and sequential representations. However, these methods still face a significant limitation in that the feature levels of model outputs and item embeddings are inconsistent, as the model's output represents high-level features while item embeddings represent low-level features, resulting in a substantial gap in feature levels, which prevents effective representation generation when computing similarity directly between them.

\subsection{Self-supervised Multi-modal Learning}

Deep neural networks have proven to be highly effective in many applications, but their performance can be limited by the availability of labeled data.  While training on large-scale labeled datasets \cite{mcauley2015image, krizhevsky2017imagenet} can improve performance, the use of deep networks can be constrained when data is scarce or obtaining annotations is challenging. 
To overcome these issues, self-supervised learning (SSL) has emerged as a promising alternative. Unlike traditional supervised learning, SSL \cite{jaiswal2020survey} enables models to learn from unlabeled data, which is often more abundant than labeled data. SSL works by training models on learning objectives derived from the training samples themselves, without the need for external annotations.

One popular approach in self-supervised learning is contrastive learning \cite{he2020momentum, chen2020simple}, where the model is trained to distinguish between a positive and a set of negative examples. This has been successfully applied in cross-modal tasks such as visual-text matching \cite{radford2021learning, xu2021videoclip, miech2019howto100m}, audio-visual alignment \cite{korbar2018cooperative, morgado2021audio}, and multi-modal combining \cite{akbari2021vatt, alayrac2020self}. These approaches have achieved state-of-the-art results in various benchmarks, demonstrating the effectiveness of self-supervised learning and contrastive learning in solving real-world problems. 

Video is an excellent example of a multimodal learning source that naturally combines multiple modalities and allows for learning from large-scale data that may not be feasible to manually annotate. Recently, there have been many efforts \cite{luo2022clip4clip, yang2021taco, xu2021videoclip} in cross-modal self-supervised learning on video. Audioclip \cite{guzhov2022audioclip} uses pre-trained models for feature extraction from three modalities: vision, audio, and text. It employs contrastive learning between each pair of modalities to align their features. FrozenInTime \cite{bain2021frozen} uses a dual encoding model to separately encode video and text, with its space-time transformer encoder capable of flexibly encoding images or videos by treating an image as a single-frame video. Wav2clip \cite{wu2022wav2clip} utilizes the CLIP \cite{radford2021learning} as the encoder for text and images, and constructs a contrastive learning task between the frozen image encoder and audio encoder to align the hidden representations of the three pre-trained modal encoders. EverythingAtOnce \cite{shvetsova2022everything} aligns different modal combinations with each other by constructing a composite contrastive learning loss between multiple modalities.

\section{Method}

In this section, we propose a self-supervised multi-modal sequential recommendation model. By leveraging self-supervised learning on multi-modal datasets, our approach effectively enhances the retrieval performance of the model, thereby improving its performance on downstream recommendation tasks. n the following, for convenience, we consider the use of two modalities: vision and text.

\subsection{Problem Statement}

\subsubsection{Input Formulation}

Assuming that the set of all items is denoted as $I$, each item $i \in I$ is represented not by a simple item ID, but by a tuple $\left(v_i, t_i\right)$ which encompasses both visual and textual information associated with the item. The visual information associated with an item is represented as $v_i = \left\{v_i^1, v_i^2, \ldots, v_i^m\right\}$, where $v_i^j$ corresponds to an image, and the length of the visual sequence denoted as $\left|v_i\right| = m$. Notably, there are two types of input information for the visual modality: videos and images. In the case of videos, the visual sequence is obtained by extracting frames from the video at regular intervals, with each $v_i^j$ representing a frame from the video. While in the case of images, the visual sequence length $\left|v_i\right| = 1$, as each item is represented by a single image.
The textual information associated with an item is denoted as $t_i = \left\{t_i^1, t_i^2, \ldots, t_i^n\right\}$, where $t_i^j$ represents a word token, and the length of the textual sequence denoted as $\left|t_i\right| = n$.

\subsubsection{Sequence Recommendation}

The objective of the sequence recommendation task is to infer a user's preferences and provide recommendations for the next item based on historical interaction data. Let us denote the user's interaction sequence, sorted chronologically, as $S=\left\{ i_1, \ldots, i_t, \ldots, i_{|S|} \right\}$, where $i_t \in I$ represents the item that the user interacted with at time step $t$, and $|S|$ denotes the length of the user's interaction sequence. Accordingly, the task objective can be formulated as predicting the probability distribution over the entire item set $I$ for the potential interaction of the user at time step $|S|+1$, which can be formulated as:
\begin{align}
p\left(i_{\left|S\right|+1}=i\vert S\right)
\end{align}

\subsection{Self-supervised Multi-modal Learning}

\begin{figure*}[h]
  \centering
  \includegraphics[width=\linewidth]{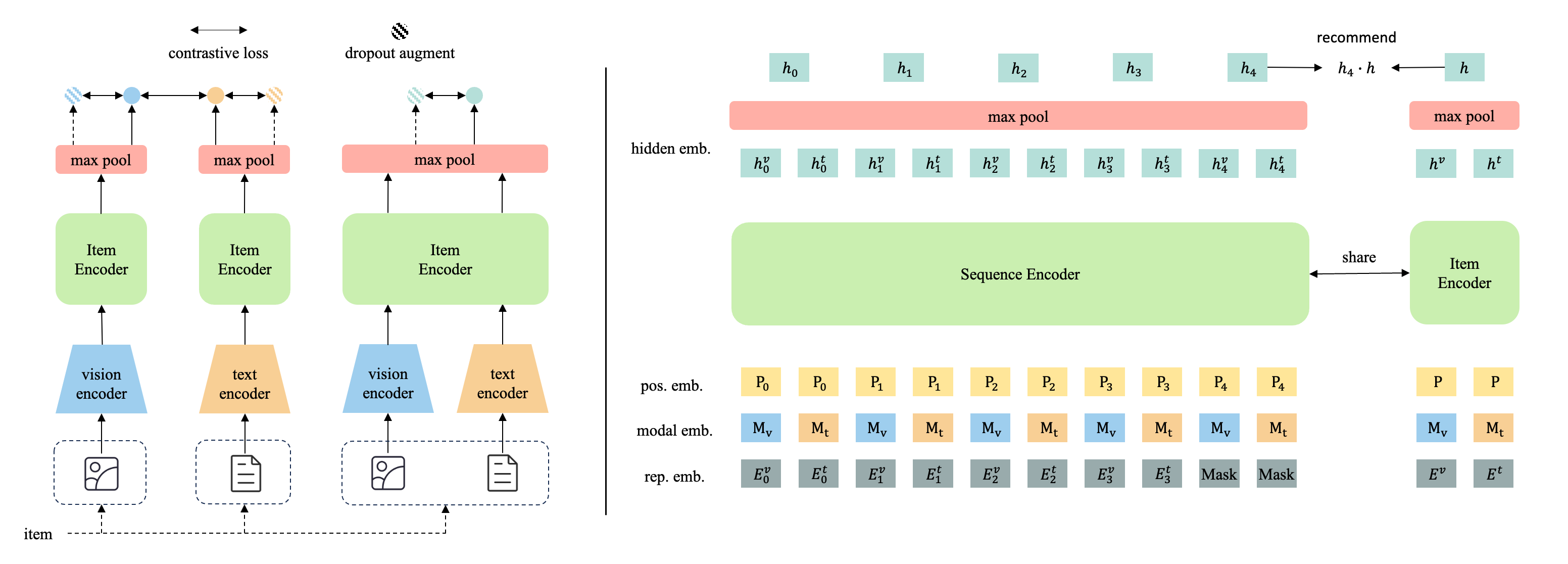}
  \caption{The framework of the self-supervised multi-modal sequential recommendation method. The left part illustrates the self-supervised multi-modal pre-training method, while the right part shows the dual-tower retrieval architecture for sequential recommendation.}
  \label{main_img}
\end{figure*}

\subsubsection{Input Representation}

In previous sequential recommendation approaches \cite{hidasi2015session, kang2018self, sun2019bert4rec}, explicit item IDs were used as trainable representations of items, which restrict the model's ability to transfer general knowledge across different domains. To overcome this limitation and enable effective transfer of pre-trained models to new recommendation scenarios, we propose to use modal information associated with items as a bridge that connects multiple domains. By extracting and mapping the modal features of items from diverse domains to a common semantic space using pre-trained modality models \cite{devlin2018bert, he2016deep, radford2021learning}, we overcome the reliance on explicit item IDs. As most recommendation scenarios involve items with both visual and textual modalities, we choose these two modalities as the input feature information for items.

For an item with visual modality, we process it as two types of input data: images and videos. For image data, we extract features directly using a pre-trained visual model (PVM) \cite{radford2021learning} to obtain the image representation. For video data, we first sample frames from the videos at regular intervals to obtain a collection of video frames, denoted as $v_i = \left\{v_i^1, v_i^2, \ldots, v_i^m\right\}$. Next, we use the PVM to extract features from all video frames, and then take the average of all frame features to obtain the representation of the video. The obtained representation then passes through a single-layer neural network, producing the visual feature embedding:
\begin{align}
E_i^v = f_{NN}( \frac{1}{|m|} \sum_{j=1}^{|m|}f_{PVM}( v_i^j ) )
\end{align}
where $E_i^v$ represents the visual feature embedding for item $i$, $f_{NN}$ denotes the neural network function, and $f_{PVM}$ represents the feature extraction function of the pre-trained visual model.

For an item with textual modality, we utilize a pre-trained language model (PLM) \cite{radford2021learning} to extract features from the associated text. Given the text token sequence $t_i = {t_i^1, t_i^2, \ldots, t_i^n}$ of item $i$, we first prepend a special token $\left[CLS\right]$ to the token sequence, and then input the concatenated sequence into the PLM. Finally, we select the token corresponding to the $\left[CLS\right]$ position in the hidden layer as the representation of the text. The obtained representation then passes through a single-layer neural network, producing the text feature embedding:
\begin{align}
E_i^t = f_{NN}(f_{PLM}( \left\{ [CLS], t_i^1, t_i^2, \ldots, t_i^n\right\} ) )
\end{align}
where $E_i^t$ represents the text feature embedding for item $i$, and $f_{PLM}$ represents the feature extraction function of the pre-trained language model.

Hence, following the feature extraction process using pre-trained modality models, an input item $i \in I$ is converted into a feature embedding tuple $(E_i^v, E_i^t)$. As the item encompasses two distinct modalities, we employ the sum of feature embedding and modality embedding as the input to the model to enhance discriminability.

\subsubsection{Multi-modal Item Encoder}

In various recommendation scenarios, items may be associated with different modalities, such as visual and textual modalities. These modalities can take the form of single modalities, denoted as $v$ or $t$, or a combination of both modalities, represented as a tuple $(v, t)$. Our objective is to develop a method that enables the model to be modality-agnostic, allowing it to map input from any modality combination of an item to a common hidden space and align them using a mapping function $f_{item}$. Considering that an item may have multiple modal inputs, the model needs to encode bidirectionally, enabling jointly embed them for representation.

Therefore, we propose a multi-modal item encoder based on the foundational structure of a transformer block \cite{vaswani2017attention}. The transformer block comprises a multi-head self-attention layer and a feed-forward network. Prior to inputting the feature embeddings of an item into the model, we add modality embeddings to them. In cases where certain modalities are absent, we replace the corresponding feature embeddings with a $\left[mask\right]$ token. Leveraging the powerful attention mechanism of the transformer block, different modalities can mutually attend to each other, facilitating information sharing and completion of missing modality information. Subsequently, we employ modality-specific projection layers to project feature embeddings into a shared hidden space, followed by normalization. Finally, a max-pooling layer is utilized to extract the salient feature information from all modalities, resulting in a unified representation of the item:
\begin{align}
h_i = \text{MaxPooling}(h_i^v, h_i^t)
\end{align}

\subsubsection{Contrastive Loss}

As our objective is to develop a modality-agnostic model that can effectively handle diverse combinations of input modalities, and align them in a shared hidden space, we draw inspiration from recent advances in contrastive learning \cite{radford2021learning, shvetsova2022everything}. Contrastive learning has shown remarkable capabilities in representation alignment and generalization, where similar input contents are encouraged to be close while inputs with distinct semantics are pushed apart. To leverage these benefits, we adopt a retrieval pre-training task based on contrastive learning. This approach enhances the semantic alignment capability among multiple modalities, addressing the challenge of inconsistent modalities associated with items in different domains. By employing contrastive learning, we enable items with similar semantics but varying modalities from distinct domains to align with each other, promoting robust and effective representation learning.

Taking text and visual modalities as an example, the model should possess the capability to align within a single modality, such as $(v, v^\prime)$ and $(t, t^\prime)$, align across modalities, such as $(v, t)$, and align modalities in combination, such as $(vt, vt^\prime)$. Here, $v^\prime$, $t^\prime$, and $vt^\prime$ represent the augmented data using an unsupervised data augmentation method \cite{gao2021simcse}. The contrastive loss function is designed as follows:

\begin{align}
\begin{aligned}
\mathcal{L} =
& \lambda_{v,v^\prime} \ell(v, v^\prime) +
\lambda_{t,t^\prime} \ell(t, t^\prime) \\
& + \lambda_{v,t} \ell(v, t) +
\lambda_{vt,vt^\prime} \ell(vt, vt^\prime)
\end{aligned}
\end{align}
where $\lambda_{x,y}$ represents the weight coefficient between modality $x$ and $y$, and $\ell(x,y)$ represents the contrastive loss using Noise Contrastive Estimation \cite{oord2018representation}:
\begin{align}
\begin{aligned}
\ell(x, y) = 
& \mathop{\mathbb{E}}\limits_{i \in |\mathcal{B}|}
\left[-log\frac{exp(sim(x_i, y_i) / \tau)}
{\begin{matrix} 
\sum_{j=1}^\mathcal{B} exp(sim(x_i, y_j) / \tau)
\end{matrix}} \right] \\
& + \mathop{\mathbb{E}}\limits_{j \in |\mathcal{B}|}
\left[-log\frac{exp(sim(y_j, x_j) / \tau)}
{\begin{matrix} 
\sum_{i=1}^\mathcal{B} exp(sim(y_j, x_i) / \tau)
\end{matrix}} \right]
\end{aligned} 
\end{align} 
where $\mathcal{B}$ represents the batch size, and $\tau$ is a temperature coefficient.

Through the utilization of contrastive loss, our model has achieved the capability to represent items in a universal manner, overcoming the limitations imposed by the size of the pretraining dataset in the recommendation domain. As a result, our model exhibits robust generalization capabilities, being able to effectively encode inputs of diverse modality combinations and align semantically similar items in the shared hidden layer space.

\subsection{Sequential Representation Learning}

Due to the sensitive nature of user interactions in recommendation domain and strict privacy regulations, there is a lack of publicly available large-scale pretraining datasets. Previous methods for sequence-based recommendation \cite{ding2021zero, hou2022towards, wang2022transrec} have relied on limited datasets for pretraining, leading to inadequate generalization capabilities and diminished performance when transferring across domains. Recognizing the intricate nature of user behavior patterns in diverse recommendation domains, our objective is to avoid introducing domain-specific biases into our model. Therefore, we learn domain-specific sequence interaction patterns for each domain.

\subsubsection{Input Representation}

In the pretraining retrieval task, as discussed in Section 3.2, the model operates on individual items without considering their relative positional relationships. However, in the downstream sequential recommendation task, the model takes into account a user's historical interaction sequence as input, where items are arranged chronologically based on interaction time, thus having explicit positional relationships. To capture this temporal order relationship, we introduce positional embeddings as an additional component of the input for the model. Specifically, for a given item i, its input embedding is obtained by summing the feature embedding, modality embedding, and positional embedding.

In this work, in order to improve the performance of our model, we utilize learnable positional embedding matrices instead of fixed sinusoid embeddings. The positional embedding matrices enable our model to capture contextual relationships and the interaction order of each item within the input sequence, leading to enhanced representations for user sequences. Moreover, considering the constraint on the maximum length $N$ of input sequences that our model can handle, when the length of an input sequence $S=\left\{ i_1, i_2, \ldots, i_{|S|} \right\}$ exceeds $N$, we truncate the sequence and retain only the last $N$ items as $S_{trunc}=\left\{ i_{|S|-N+1}, \ldots, i_{|S|} \right\}$.

\subsubsection{Multi-modal Sequence Encoder}

The sequence encoder serves two primary objectives: firstly, to encode each item based on its multiple modal information, and secondly, to enhance the contextual representation of each item by incorporating the interaction relationships among items within the sequence. As the item encoder has already acquired generalized item representations during the pre-training phase, the sequence encoder leverages the model parameters of the item encoder directly and focuses on learning the interaction relationships within the sequence. The structure of the sequence encoder closely resembles that of the item encoder, employing transformer blocks which are widely used in various applications. The attention mechanism, known for its robust performance, enables each item in the sequence to attend to contextual information, facilitating the model in inferring representations of each item based on the contextual cues within the sequence.

\subsubsection{Masked Item Prediction}

The primary objective of sequence recommendation is to infer user preferences and provide recommendations for the next potential item based on historical interaction information. To achieve this, the sequential encoder needs to possess the ability to predict items by leveraging contextual content information. In light of the Masked Language Model task \cite{devlin2018bert}, we propose a Masked Item Prediction task for our model.

In the Masked Item Prediction task, for a given input sequence $S=\left\{ i_1, i_2, \ldots, i_{|S|} \right\}$, each item $i$ is replaced with a special token $[mask]$ with a probability $p$. The model is then tasked with predicting the original item based on its contextual content information. As our model represents each item with multiple modal representation embeddings as inputs, for the masked item, we replace all its associated modal representation embeddings with the token $[mask]$.

During testing, the objective of predicting the next potential interaction item requires the addition of the token $[mask]$ at the end of the user's historical interaction behavior sequence. This enables the model to predict the next item in the sequence that the user is likely to interact with.

To ensure consistency in input between training and testing, where the token $[mask]$ only appears at the last position during testing, we adopt a specific strategy during training. The last position of the input sequence sample is always replaced with the token $[mask]$ during training. For other positions of the token $[mask]$ in the sequence, we employ three replacement strategies: (1) replacing the item with the $[mask]$ token 80\% of the time, (2) replacing the item with a randomly selected item 10\% of the time, and (3) keeping the item unchanged 10\% of the time. This approach addresses the inconsistency in input between training and testing, ensuring that the model is trained to effectively predict the masked items in the sequence during testing.

\section{Experiment}

\subsection{Pre-training Experiments}

\subsubsection{Datasets}

During the pre-training phase, we employed the WebVid dataset \cite{bain2021frozen} as our training data. This dataset comprises a large-scale collection of video-text pairs, totaling 10 million pairs obtained from stock footage websites. Videos, being a rich source of diverse modal features, are particularly well-suited for our self-supervised multi-modal pre-training task. To simplify the training process, we utilized the pre-extracted features \footnote{https://huggingface.co/datasets/iejMac/CLIP-WebVid} obtained from CLIP (ViT-B/32) at a frame rate of 1FPS.

In addition, we employed MSR-VTT \cite{xu2016msr} as our test set to assess the generalization ability of our model. MSR-VTT consists of 10,000 videos, each ranging in length from 10 to 32 seconds, and a total of 200,000 captions. For evaluation purposes, we utilized a set of 1,000 test clips to assess the performance of our model.

\subsubsection{Evaluation Metrics}

We employed the standard Recall@K retrieval metric to evaluate the performance of our model. Recall@K assesses the proportion of test samples in which the correct result is present among the top K retrieved points for a given query sample. We report the results for Recall@5 and Recall@10 as performance indicators for our model.

\subsubsection{Implement Details}

We utilize CLIP (ViT-B/32) as the visual encoder and textual encoder to extract modal features. For MSR-VTT, we evenly partition the videos into 10 clips based on their total length, and extract one frame image from each clip for feature extraction. During training, the parameters of these modal feature encoders are kept frozen, and only the item encoder is trained.

The model parameters are optimized using the AdamW optimizer with the learning rate of 5e-5, and exponential decay is applied to the learning rate with a decay rate of 0.9. We set the maximum number of video frames allowed in the model to 10 and the maximum length of text tokens to 77. The batch size is configured to 48000 for training. The embedding dimension of the model is 512, with 2 layers and 8 heads. We set the embedding dropout and hidden dropout to 0.2 and 0.5, respectively. Additionally, the weight coefficients $\lambda_{v,v^\prime}$, $\lambda_{t,t^\prime}$, $\lambda_{v,t}$, and $\lambda_{vt,vt^\prime}$ are all 0.25, and the temperature coefficient is 0.05. The model is pretrained for 15 epochs using 8 V100 GPUs (32GB memory), and the training process is completed in 5 hours.

\subsubsection{Performance}

The objective of evaluating the pre-trained model is to assess its efficacy in aligning diverse modalities and its ability to generalize. As the downstream sequential recommendation task involves predicting representations of specific item positions and retrieving items based on these representations, the retrieval performance of the pre-trained model directly impacts its performance on the downstream task. To evaluate the generalization ability of the model, zero-shot testing is conducted directly on the test set. To assess the effectiveness of the alignment between different modalities, the text-to-video retrieval task is chosen as the evaluation metric.

Table \ref{pre-train-result} presents the results of our model compared to Clip, EverythingAtOnce, and FrozenInTime on the zero-shot text-to-video retrieval task using MSR-VTT. The table reveals that our model outperforms Clip in terms of retrieval performance. By integrating the features of multiple modalities through the item encoder, our model exhibits improved representation capability for items compared to using Clip directly for extracting multi-modality features. Furthermore, when comparing our model with EverythingAtOnce, despite both models utilizing the same backbone, the variance in pre-training datasets also yields significant performance differences on the test set. Hence, training the model on larger pre-training datasets can further augment its generalization ability and enhance its performance.

\subsection{Downstream Task Experiments}

\begin{table}
  \caption{Results of pre-training on MSR-VTT dataset for zero-shot text-to-video retrieval. "R@k" is short for "Recall@k".}
  \label{pre-train-result}
  \small
  \begin{tabular}{l|c|c|c|cc}
    \toprule
    Method & \makecell{Pretrain \\ Dataset} & Backbone & \makecell{Backbone \\ Trainable} & R@5 & R@10 \\
    \midrule
    Clip\cite{radford2021learning} & - & ViT-B/32 & \ding{53} & 42.1 & 51.7 \\
    EAO\cite{shvetsova2022everything} & HW100M & Clip (ViT-B/32) & \ding{53} & 32.5 & 42.4 \\
    FIT\cite{bain2021frozen} & CC+WV & Clip (ViT-B/32) & \ding{51} & 46.9 & 57.2 \\
    Ours & WV10M & Clip (ViT-B/32) & \ding{53} & 47.3 & 59.7 \\
  \bottomrule
\end{tabular}
\end{table}

\begin{table}
  \caption{Statistics of the downstream task datasets after preprocessing}
  \label{downstrean-dataset}
  \begin{tabular}{l|rrrr}
    \toprule
    Dataset     & \#Users   & \#Items   & \#Actions & Sparsity  \\
    \midrule
    Beauty      & 22,363    & 12,101    & 198,502   & 99.93\%   \\
    Sports      & 35,598    & 18,357    & 296,337   & 99.95\%   \\
    Clothing    & 39,387    & 23,033    & 278,677   & 99.97\%   \\
    Home        & 66,519    & 28,237    & 551,682   & 99.97\%   \\
    ML-1m       & 6,040     & 3,416     & 999,611   & 95.16\%   \\
  \bottomrule
\end{tabular}
\end{table}

\begin{table*}
  \caption{Performance comparison of different recommendation models. The best and the second-best performances are denoted in bold and underlined fonts, respectively. “Improv.” indicates the relative improvement ratios of the proposed approach over the best performance baselines. The features used for item representations of each compared model have been listed, whether ID, feature (F), or both (ID+F).}
  \label{downstream-result}
  \begin{tabular}{llccccccccr}
    \toprule
    \textbf{Dataset}               & \textbf{Metric}    & 
    \textbf{Pop} & \textbf{GRU4Rec$_{ID}$}     & \textbf{SASRec$_{ID}$}    & \textbf{FDSA$_{ID+F}$}      &
    \textbf{S$^3$-Rec$_{ID+F}$}   & \textbf{DuoRec$_{ID}$}    & \textbf{UniSRec$_F$}   & \textbf{Ours$_F$}      & Improv      \\
    \midrule
    \multirow{4}{*}{Beauty} & Recall@10 & 
    0.0189 & 0.0627      & 0.0842    & \underline{0.0874} &
    0.0862      & 0.0853    & 0.0743    & \textbf{0.0949}    & 8.58\%      \\
                            & Recall@50 &
    0.0555 & 0.1498      & 0.1766    & 0.1838    &
    0.1871      & 0.1854    & \underline{0.1885}    & \textbf{0.2302}         & 22.12\%               \\
                            & NDCG@10   &
    0.0093 & 0.0330      & 0.0416    & \underline{0.0462}    & 
    0.0434      & 0.0449    & 0.0351    & \textbf{0.0476} & 3.03\%     \\    
                            & NDCG@50   &
    0.0174 & 0.0520      & 0.0618    & \underline{0.0680} &
    0.0653      & 0.0663    & 0.0599    & \textbf{0.0754}    & 10.88\%     \\    
    \midrule
    \multirow{4}{*}{Sports} & Recall@10 & 
    0.0177 & 0.0351      & 0.0477    & 0.0504    &
    \underline{0.0523}      & 0.0514    & 0.0513    & \textbf{0.0635}         & 21.41\%   \\
                            & Recall@50 &
    0.0510 & 0.0928      & 0.1114    & 0.1183    & 
    0.1210      & 0.1176    & \underline{0.1314}    & \textbf{0.1607}         & 22.30\%   \\    
                            & NDCG@10   &
    0.0097 & 0.0183      & 0.0222    & \underline{0.0276}    &
    0.0250      & 0.0248    & 0.0252    & \textbf{0.0323}    & 17.03\%     \\    
                            & NDCG@50   &
    0.0169 & 0.0308      & 0.0360    & 0.0422    &
    0.0399      & 0.0391    & \underline{0.0425}    & \textbf{0.0534}         & 25.65\%   \\   
    \midrule
    \multirow{4}{*}{Clothing} & Recall@10 & 
    0.0085 & 0.0158      & 0.0268    & 0.0283    &
    0.0370      & 0.0313    & \underline{0.0382}    & \textbf{0.0452}         & 18.32\%   \\
                            & Recall@50 &
    0.0303 & 0.0477      & 0.0608    & 0.0682    & 
    0.0858      & 0.0677    & \underline{0.1021}    & \textbf{0.1269}         & 24.29\%   \\    
                            & NDCG@10   &
    0.0045 & 0.0078      & 0.0122    & 0.0156    &
    0.0169      & 0.0148    & \underline{0.0191}    & \textbf{0.0218}         & 14.14\%   \\    
                            & NDCG@50   &
    0.0092 & 0.0146      & 0.0195    & 0.0242    &
    0.0275      & 0.0227    & \underline{0.0328}    & \textbf{0.0395}         & 20.43\%   \\  
    \midrule
    \multirow{4}{*}{Home}   & Recall@10 & 
    0.0136 & 0.0189      & 0.0303    & 0.0277    &
    \underline{0.0329}      & 0.0309    & 0.0265    & \textbf{0.0388}         & 17.93\%   \\
                            & Recall@50 &
    0.0436 & 0.0532      & 0.0638    & 0.0680    &
    0.0690      & 0.0669    & \underline{0.0725}    & \textbf{0.0969}         & 33.66\%   \\    
                            & NDCG@10   &
    0.0069 & 0.0098      & 0.0152    & 0.0155    &
    \underline{0.0163}      & 0.0156    & 0.0135    & \textbf{0.0195}         & 19.63\%   \\    
                            & NDCG@50   &
    0.0134 & 0.0171      & 0.0224    & \underline{0.0242}    &
    0.0241      & 0.0234    & 0.0234    & \textbf{0.0321}    & 32.64\%     \\  
    \midrule
    \multirow{4}{*}{ML-1m}  & Recall@10 & 
    0.0749 & 0.2988      & 0.2993    & \underline{0.3028}    &
    0.3002      & 0.3013    & 0.1472    & \textbf{0.3124}    & 3.17\%      \\
                            & Recall@50 &
    0.2110 & 0.5412      & 0.5457    & \underline{0.5523}    &
    0.5467      & 0.5421    & 0.4114    & \textbf{0.5650}    & 2.30\%      \\    
                            & NDCG@10   &
    0.1310 & 0.1731      & 0.1690    & \underline{0.1744}    &
    0.1694      & 0.1697    & 0.0665    & \textbf{0.1835}    & 5.22\%      \\    
                            & NDCG@50   &
    0.1522 & 0.2264      & 0.2237    & \underline{0.2296}    &
    0.2240      & 0.2231    & 0.1242    & \textbf{0.2392}    & 4.18\%      \\  
   \bottomrule
  \end{tabular}
\end{table*}

\subsubsection{Datasets}

In order to evaluate the effectiveness of our proposed model, we selected five open-source datasets from real-world platforms, taking into consideration the total number of users, items, actions, and dataset sparsity. Among these datasets, four are Amazon platform  datasets \footnote{http://jmcauley.ucsd.edu/data/amazon/} \cite{mcauley2015image}, including "Beauty," "Sports and Outdoors," "Clothing Shoes and Jewelry," and "Home and Kitchen." Additionally, to evaluate the generalization ability of the model across different platforms, we also selected Movielens-1M \footnote{https://grouplens.org/datasets/movielens/1m/} \cite{harper2015movielens}, which is widely used for evaluating recommendation algorithms.

We followed the approach used in previous works \cite{kang2018self, sun2019bert4rec, qiu2022contrastive, zhou2020s3} to process these datasets, by keeping the five-core datasets and filtering out users and items with fewer than five interactions. Subsequently, we grouped the interactions by users and sorted them in ascending order based on the timestamps. For the four Amazon datasets, we crawled the image links associated with the Amazon dataset items as visual modality information. Items without image links were labeled as having missing visual modality information. Furthermore, we concatenated the titles and descriptions associated with the Amazon dataset items as textual modality information. For Movielens-1M, we crawled corresponding movie trailers from YouTube \footnote{https://www.youtube.com/} based on the movie names in the dataset as visual modality information, and concatenated the movie names and tags as textual modality information. The statistical information of the preprocessed datasets is shown in Table \ref{downstrean-dataset}.

\subsubsection{Evaluation Metrics}

To evaluate the performance of models, we utilize the top-k Recall and top-k Normalized Discounted Cumulative Gain (NDCG) metrics, which are commonly used in related works \cite{hou2022towards, xie2022contrastive, sun2019bert4rec}. Recall measures the presence of the positive item, while NDCG takes into account both the rank position and the presence. In our experiments, we report Recall and NDCG at k = 10, 50.
In addition, we adopt the leave-one-out strategy, which has been widely employed in previous works \cite{zhou2020s3, qiu2022contrastive, wang2022transrec}. Specifically, for each user, we retain the last interaction item as the test data, and the item just before the last as the validation data. The remaining items are used for training. We rank the ground-truth item of each sequence against all other items for evaluation on the test set, and finally calculate the average score across all test users.

\subsubsection{Baselines}
We compare the proposed approach with the following baseline methods:

\begin{itemize}
    \item \textbf{Pop} A non-personalized approach involves recommending the same items to all users. These items are determined based on their popularity, measured by the highest number of interactions across the entire set of items.
    \item \textbf{GRU4Rec} \cite{hidasi2015session} proposes an approach for session-based recommendations using recurrent neural networks with Gated Recurrent Units.
    \item \textbf{SASRec} \cite{kang2018self} proposes a self-attention based sequential model which uses the multi-head attention mechanism to recommend the next item.
    \item \textbf{FDSA} \cite{zhang2019feature} integrates various heterogeneous features of items into feature sequences with different weights through a vanilla attention mechanism.
    \item \textbf{S$^3$-Rec} \cite{zhou2020s3} is a self-supervised learning approach for sequential recommendation that utilizes mutual information maximization (MIM) to learn correlations among attribute, item, subsequence, and sequence.
    \item \textbf{DuoRec} \footnote{https://github.com/RuihongQiu/DuoRec}  \cite{qiu2022contrastive} addresses the representation degeneration problem in sequential recommendation. It uses contrastive regularization to reshape the distribution of sequence representations and improve the item embeddings distribution.
    \item \textbf{UniSRec} \footnote{https://github.com/RUCAIBox/UniSRec} \cite{hou2022towards} utilizes the associated description text of items to learn transferable representations across different recommendation scenarios.
\end{itemize}

\subsubsection{Implement Details}

For DuoRec and UniSRec, we utilized the source code provided by their respective authors. For the other methods, we implemented them using RecBole \cite{zhao2021recbole}, a widely used open-source recommendation library. All hyper-parameters were set based on the recommendations from the original papers. Additionally, fine-tuning of all baseline models was conducted on the five downstream recommendation datasets.

For our proposed model, we use the AdamW optimizer with a learning rate of 1e-3 and configure the batch size to 8192. The mask item ratio of the model is 0.2. For the four Amazon datasets, we set the maximum sequence length to 20, the embedding dropout to 0.2 and the hidden dropout to 0.5. For Movielens-1M, we set the maximum sequence length to 100, the embedding dropout to 0.2 and the hidden dropout to 0.2. Furthermore, we adopted early stopping with a patience of 10 epochs to prevent overfitting and set Recall@10 as the indicator.

\subsubsection{Performance}

We conducted a comparative analysis of the proposed approach against several baseline methods on five publicly available datasets, and the results of our experiments are presented in Table \ref{downstream-result}.

Based on the experimental results, we observed that models which incorporate both modal features and ID embeddings, such as FDSA and S$^3$-Rec, tend to outperform models that solely rely on ID embeddings, such as SASRec and DuoRec. This can be attributed to the fact that these models not only leverage trainable ID embeddings but also utilize modal features of items as supplementary information, which provides additional cues during the inference process. Notably, UniSRec achieves competitive performance with these modal-assisted models on the four Amazon datasets by exclusively utilizing text features as item representations without employing ID embeddings, largely due to its pretraining on five Amazon datasets \footnote{Amazon Review Dataset: "Grocery and Gourmet Food", "Home and Kitchen", "CDs and Vinyl", "Kindle Store" and "Movies and TV"}. However, as evident from the table, UniSRec exhibits inferior performance on Movielens-1M compared to other methods. This can be attributed to the fact that UniSRec was only pretrained on a relatively small-scale dataset, resulting in limited generalization capability and inadequate transfer of generic knowledge to cross-platform datasets.

Lastly, our proposed method was systematically compared against all baseline models, and the results clearly indicate that our model achieves superior performance on all datasets. Notably, our model exhibits significant improvements on both sparse datasets (e.g., Amazon review dataset) and dense datasets (e.g., Movielens-1M), outperforming all other baseline models by a substantial margin on sparse datasets in particular. Unlike these baselines, our model does not rely on item embedding-based approaches for sequential recommendation. Instead, it adopts a retrieval-based approach facilitated by self-supervised multi-modal pre-training task, enabling it to learn universal item representations and enhance its generalization capability. In the downstream sequential recommendation task, our model jointly encodes sequences and items, and predicts tokens at masked positions in the sequences using the Masked Item Prediction task for item retrieval. The final results conclusively demonstrate the superior recommendation performance of our proposed method on datasets from diverse domains and varying levels of sparsity.

\subsection{Further Analysis}

\subsubsection{Sparsity Influence}

\begin{table}
  \caption{Statistical Information of Sports and Home datasets under Different k-core Strategies.}
  \label{sparsity-experiment}
  \begin{tabular}{lccccr}
    \toprule
    Dataset                 & k-core                    & 
    \#Users                 & \#Items                   & 
    \#Actions               & Sparsity                  \\
    \midrule
    \multirow{4}{*}{Sports}& 
      4 & 74,224 & 33,980 & 499,812 & 99.98\%  \\
    & 5 & 35,598 & 18,357 & 296,337 & 99.95\% \\
    & 6 & 17,281 & 9,758 & 169,846 & 99.90\% \\
    & 7 & 8,389 & 5,135 & 93,449 & 99.78\% \\
    \midrule
    \multirow{4}{*}{Home}& 
      4 & 128,156 & 46,325 & 855,824 & 99.99\%  \\
    & 5 & 66,519 & 28,237 & 551,682 & 99.97\% \\
    & 6 & 35,452 & 17,382 & 351,703 & 99.94\% \\
    & 7 & 18,748 & 10,606 & 217,497 & 99.89\% \\
   \bottomrule
  \end{tabular}
\end{table}

\begin{figure}
  \centering
  \includegraphics[width=\linewidth]{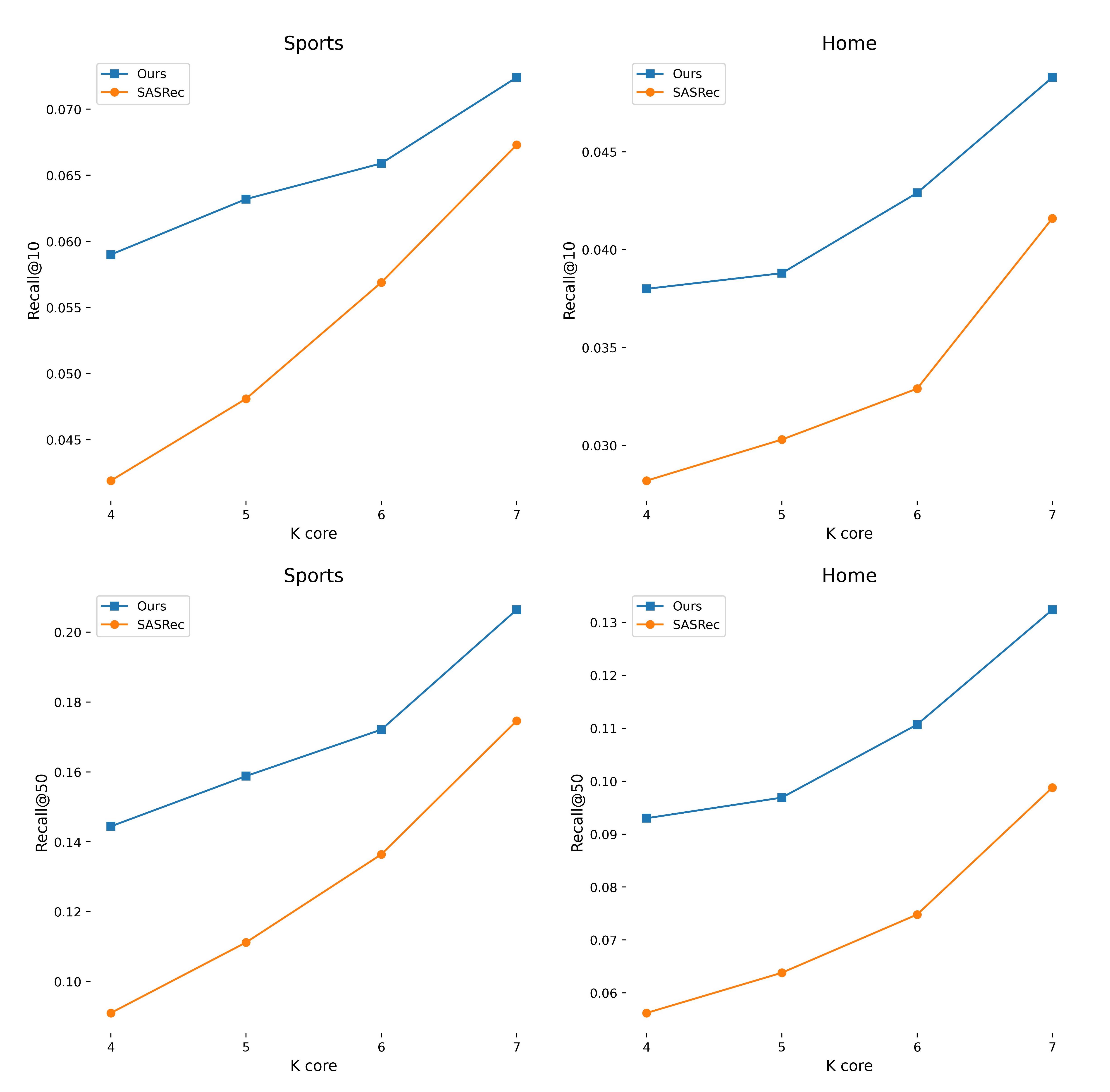}
  \caption{Model Performance across Different Dataset Sparsity Levels.}
  \label{sparsity-img}
\end{figure}

The experimental results from Table \ref{downstream-result} suggest that the effectiveness of our model is closely related to the sparsity of datasets. Sparsity, defined as the proportion of missing values in relation to the total values in a dataset, is a critical measure in recommendation systems. Higher sparsity indicates a larger number of missing values in the dataset, while lower sparsity indicates fewer missing values. In the context of recommendation systems, the user-item rating matrix is typically sparse, as users only rate a small subset of all items.
However, existing methods \cite{kang2018self, sun2019bert4rec, qiu2022contrastive, zhou2020s3} have commonly adopted a direct approach of retaining the five-core datasets and filtering out users and items with fewer than five interactions, without thoroughly investigating the potential impact of different k-core filtering strategies on model performance. To address this research gap, we conducted experiments on the Sports and Home datasets using various k-core filtering strategies and evaluated the model's performance under different filtering strategies. The statistical information of the Sports and Home datasets under different k-core filtering strategies is presented in Table \ref{sparsity-experiment}. The table reveals that as the k-core decreases, the sparsity of these datasets increases. Specifically, when the k-core value decreases from 7 to 4, the sparsity of the Sports dataset increases from 99.78\% to 99.98\%, and the sparsity of the Home dataset increases from 99.89\% to 99.99\%.

The experimental results of our model on datasets with different levels of sparsity are presented in Figure \ref{sparsity-img}. We compared our model with a baseline model, SASRec, in the experiments. As shown in the figure, the evaluation results of both our model and SASRec significantly decrease as k-core decreases. This is due to the fact that as the sparsity of the dataset increases, there are a large number of users and items with only a few interactions, which is a common cold start problem in the recommendation field. However, the performance degradation of our model is noticeably smaller than that of SASRec as k-core decreases. The average improvement of our model compared to SASRec on two datasets, in terms of the Recall@10 metric, increased from 12.45\% at k-core=7 to 37.78\% at k-core=4. Similarly, the average improvement in terms of the Recall@50 metric increased from 26.11\% at k-core=7 to 62.08\% at k-core=4. Therefore, the experimental results further confirm that our model is capable of effectively addressing the cold start problem in the recommendation field.

\subsubsection{Ablation Study}

\begin{figure}
  \centering
  \includegraphics[width=\linewidth]{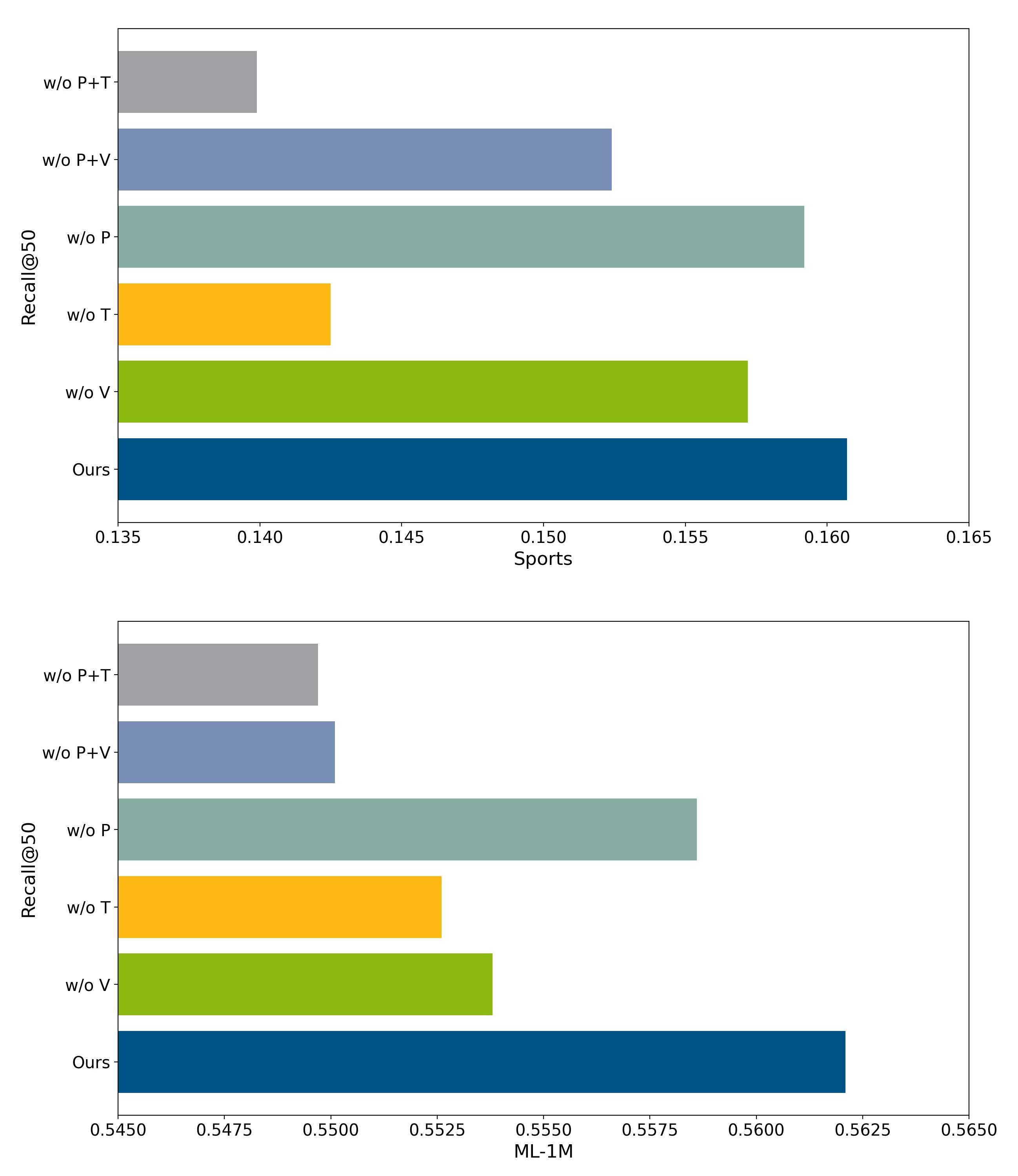}
  \caption{Ablation study of our model on Sports and Movielens-1M.}
  \label{ablation-img}
\end{figure}

In this section, we will conduct an in-depth analysis of the impact of each proposed technique and component on the performance of our system. To this end, we compare our proposed model with several variants, including:
(1) \underline{$w/o$ $V$}: without the visual modality.
(2) \underline{$w/o$ $T$}: without the text modality.
(3) \underline{$w/o$ $P$}: without pre-training.
(4) \underline{$w/o$ $P+V$}: without pre-training and the visual modality.
(5) \underline{$w/o$ $P+T$}: without pre-training and the text modality.

The results of the abovementioned experiments are presented in Figure \ref{ablation-img}. Based on the experimental results, it can be inferred that our model architecture effectively integrates features from multiple modalities, and the incorporation of both visual and text modalities leads to superior performance compared to using them individually. Furthermore, the comparison between pre-trained and non-pre-trained models reveals that regardless of whether the system has access to both visual and textual modalities or only one modality, pre-training consistently enhances the performance of the model in downstream recommendation tasks. This further corroborates the notion that our pre-training task adeptly learns representation alignment between diverse modalities and promotes mutual reinforcement among them.

\subsubsection{Backbone}

\begin{table}
  \caption{Evaluation of Different Text Backbones on the Sports. $t$ is for text modality only. "R@k" is short for "Recall@k", and "N@k" is short for "NDCG@k".}
  \label{backbone-result}
  \small
  \begin{tabular}{l|c|cccc}
    \toprule
    \multirow{2}{*}{Method} & \multirow{2}{*}{Backbone} & 
    \multicolumn{4}{c}{Metrics} \\
    & & R@10 & R@50 & N@10 & N@50 \\
    \midrule
    UniSRec$_t$ & Bert (base) & 0.0513 & 0.1314 & 0.0252 & 0.0425 \\
    \midrule
    Our$_t$ & Bert (base) & 0.0588 & 0.1501 & 0.0299 & 0.0497 \\
    Our$_t$ & Clip (ViT-B/32) & 0.0603 & 0.1524 & 0.0309 & 0.0501 \\
   \bottomrule
  \end{tabular}
\end{table}

We additionally employed other text backbones to evaluate our model. Specifically, we utilized the Bert as our text encoder instead of the Clip. We compared the performance of our model with UniSRec$_t$ on the Sports, and the results are presented in Table \ref{backbone-result}. Since we did not use any pretraining data to extract features with the Bert, the results of our model are without pertaining.

From the table, it can be observed that when using the Bert as the backbone, our model outperforms the UniSRec$_t$ in all four evaluation metrics, with an average improvement of 16.11\%. This demonstrates the effectiveness of our proposed model architecture. Furthermore, by replacing the Bert with the Clip, our model shows an average improvement of 2.06\%. This further confirms that utilizing a more powerful feature extractor can lead to additional performance gains in our model.

\section{Conclutions}

In this paper, we propose a self-supervised multi-modal sequential recommendation method. In contrast to conventional sequential recommendation methods that rely on explicit item IDs, our approach leverages the feature information associated with items to represent them. To address the issue of inconsistent feature levels between the model output and item embeddings, we introduce a dual-tower retrieval architecture for sequential recommendation. In this architecture, the predicted embedding from the user encoder is used to retrieve the generated embedding from the item encoder, thereby mitigating the problem of inconsistent feature levels. Additionally, we share parameters between the two encoders to facilitate mutual reinforcement of information. To further enhance the retrieval performance of the model, we also propose a self-supervised multi-modal pre-training approach. By constructing contrastive learning tasks among different feature combinations of items, our model is capable of improving its fine-grained discrimination ability for items and learning alignment of different modality features in the latent space. Extensive experiments conducted on five publicly available datasets demonstrate the effectiveness of our proposed model.

\bibliographystyle{ACM-Reference-Format}
\bibliography{acmart}

\end{document}